\documentclass[chicago,numberedappendix,tighten]{emulateapj}

\usepackage{natbib}

\usepackage{graphicx}
\usepackage[space]{grffile}
\usepackage{latexsym}
\usepackage{amsfonts,amsmath,amssymb}
\usepackage{url}
\usepackage[utf8]{inputenc}
\usepackage{fancyref}
\usepackage{hyperref}
\hypersetup{colorlinks=false,pdfborder={0 0 0},}
\usepackage{textcomp}
\usepackage{longtable}
\usepackage{booktabs}

\shorttitle{Young Remnants of Type Ia Supernovae}
\shortauthors{Chakraborti et al.}

\begin{document}


\title{Young Remnants of Type Ia Supernovae and Their Progenitors:\\A Study of SNR G1.9+0.3}


\author{Sayan Chakraborti\altaffilmark{1}}
\author{Francesca Childs\altaffilmark{2}}
\author{Alicia Soderberg\altaffilmark{3}}
\affil{Institute for Theory and Computation, Harvard-Smithsonian Center for Astrophysics, 60 Garden Street, Cambridge, MA 02138, USA}

\altaffiltext{1}{Society of Fellows, Harvard University, Cambridge, MA 02138, USA}
\altaffiltext{2}{Harvard College, Harvard University, Cambridge, MA 02138}
\altaffiltext{3}{Department of Astronomy, Harvard University, Cambridge, MA 02138, USA}

\email{schakraborti@fas.harvard.edu}




\begin{abstract}
Type Ia supernovae, with their remarkably homogeneous light curves and
spectra, have been used as standardizable candles to measure the
accelerating expansion of the Universe. Yet, their progenitors remain
elusive. Common explanations invoke a degenerate star (white dwarf)
which explodes upon reaching close to the Chandrasekhar limit, by either
steadily accreting mass from a companion star or violently merging with
another degenerate star. We show that circumstellar interaction in young
Galactic supernova remnants can be used to distinguish between these
single and double degenerate progenitor scenarios. Here we propose
a new diagnostic, the Surface Brightness Index, which can be computed
from theory and compared with Chandra and VLA observations. We use this
method to demonstrate that a double degenerate progenitor can explain the
decades-long flux rise and size increase of the youngest known Galactic SNR
G1.9+0.3. We disfavor a single degenerate scenario. We attribute the
observed properties to the interaction between a steep ejecta profile
and a constant density environment. We suggest using the upgraded VLA
to detect circumstellar interaction in the remnants of historical
Type Ia supernovae in the Local Group of galaxies. This may settle the
long-standing debate over their progenitors.
\end{abstract}

\bibliographystyle{hapj}



\keywords{ISM: supernova remnants --- radio continuum: general --- X-rays: general
--- binaries: general --- circumstellar matter --- supernovae: general
--- ISM: individual objects(\objectname{SNR G1.9+0.3})}


\section{Introduction} 
Type I supernovae were classified by \citet{1941PASP...53..224M} to be
a largely homogeneous group characterized by the lack of Hydrogen in their
spectra. A major subset called Type Ia, which have early-time spectra
with strong Si II \citep{1997ARA&A..35..309F}, are believed to come from
the thermonuclear explosions of degenerate stellar cores \citep{1990RPPh...53.1467W}.
Despite their homogeneity, peak absolute magnitudes of Type Ia supernovae are
not constant. \citet{1993ApJ...413L.105P} related their peak brightness to the
width of their light curve. This allowed \citet{1996ApJ...473...88R} to standardize
them as reliable distance indicators. Type Ia supernovae have
been used as standard candles, leading to the discovery of the
accelerating expansion of the
Universe \citep{1998AJ....116.1009R,1998ApJ...507...46S,1999ApJ...517..565P}.
As a consequence of their importance in astronomy and cosmology, Type Ia
supernovae are the subject of various theoretical and observational studies.
Yet, much remains to be known of the stellar systems that produce these
explosions.

It is generally agreed upon \citep{2000ARA&A..38..191H} that Type Ia supernovae
mark catastrophic explosions of white dwarfs near and or above the
\citet{1931ApJ....74...81C} limit.
Accreted mass, required to destabilize the white dwarf, is transfered from a binary
companion whose nature is as yet unknown.
The single degenerate (SD) model \citep{1973ApJ...186.1007W,1982ApJ...253..798N} uses
a progenitor system with a white dwarf and a non-degenerate companion. The companion
can be a main-sequence, sub-giant, He star, or
red-giant. In contrast, the double degenerate (DD)
model \citep{1984ApJ...277..355W,1984ApJS...54..335I} relies on
the merging of two white dwarfs.

Red supergiant progenitors of Type IIP supernovae \citep{2009MNRAS.395.1409S} have
been identified in pre-explosion images of host galaxies. \citet{2006ApJ...641.1029C}
suggested using the interaction of the supernova ejecta with the circumstellar
matter as a probe of mass loss from these red supergiants. Circumstellar interaction is now
being used to constrain the nature of Type IIP supernova
progenitors \citep{2012ApJ...761..100C,2013ApJ...774...30C,2015arXiv151006025C}.

Unlike the massive stellar progenitors of core collapse supernovae, many of the
putative progenitors for Type Ia supernovae are too faint to be
detected in external galaxies through direct imaging with the present generation
of optical telescopes. As a result circumstellar interaction, or lack
thereof, is a promising method for discerning the progenitors of Type Ia supernovae.
Properties of supernova remnants may provide a consistency check 
for the models of Type Ia supernova progenitors \citep{2007ApJ...662..472B}.
Circumstellar interaction in Type I supernovae
may produce radio emission \citep{1984ApJ...285L..63C}.
Recently, the lack of such early radio \citep{2012ApJ...750..164C,2012ApJ...746...21H}
and X-ray \citep{2012ApJ...751..134M} emission, from Type Ia SN 2011fe, has
ruled out much parameter space for SD scenarios. We seek to extend the scope of
such studies to young supernova remnants.

We predict temporal evolution of size and optically thin
radio and X-ray synchrotron emission from a young supernova remnant. Radio and
X-ray lightcurve during this regime shows
distinctly different behavior for SD and DD scenarios. We compare
these with long term radio and X-ray observations of the youngest known Galactic
supernova remnant, SNR G1.9+0.3. We develop a diagnostic, the Surface Brightness
Index, to compare the flux and size evolution. We show that observations are inconsistent
with a SD scenario and can be explained by a DD scenario. So we favor a scenario in
which two degenerate stars collided in a nearly constant density environment to produce
the supernova that made SNR G1.9+0.3. We suggest that the way to discern the progenitor 
systems of Type Ia supernova remnants is to measure the change in radius and flux over
time and then compare them with our models to check which one is favored. We also
show that two separate spectral indices should be expected for the X-ray and Radio
emission, when considering electron cooling due to synchrotron losses. We find that
this effect is also observed in SNR G1.9+0.3.

\section{SNR G1.9+0.3}
Radio surveys, using the Very Large Array (VLA), identified SNR G1.9+0.3 as the
smallest ($R\sim2$ pc), and therefore possibly the youngest Galactic supernova
remnant \citep{1984Natur.312..527G}.
Chandra X-ray Observatory data confirmed that this young remnant is in
the freely expanding phase as a X-ray-synchrotron-dominated shell supernova
remnant \citep{2008ApJ...680L..41R}. Subsequent radio and X-ray observations confirmed
its expansion and brightening \citep{2008MNRAS.387L..54G, 2014ApJ...790L..18B}. Spectral
variations in X-rays, interpreted in terms of magnetic field
obliquity dependence of cosmic ray acceleration, have been used
to argue for a Type Ia event \citep{2009ApJ...695L.149R}.
Ejecta distribution asymmetry and inhomogeneous abundances have also
been interpreted in context of Type Ia models \citep{2013ApJ...771L...9B}.
Furthermore, the remnant is not associated with any known star-forming region.
All these point towards SNR G1.9+0.3 being a young remnant
of a thermonuclear supernova; a Galactic Type Ia supernova in the 19th century,
unobserved due to the large extinction along the Galactic plane. In this
paper we develop a method to discern the progenitor systems of Type Ia remnants
and demonstrate it using SNR G1.9+0.3.

\section{Circumstellar Interaction} \label{circum_interact}
Most early emission from supernovae is powered by heating due to
radioactive decay. Whereas, most emission from old supernova remnants
is powered by cooling of shock heated ejecta and circumstellar matter.
Less attention is given to late emission from supernovae
and young remnants where radioactive heating becomes less important and is
gradually overtaken by circumstellar interaction. SNR G1.9+0.3 provides
a unique window into this young remnant stage, where circumstellar interaction
is the major source of heating, yet the swept up mass is low enough that
the remnant is in nearly free expansion. This allows us to build a simple model
for radio synchrotron emission from a young supernova remnant.

\begin{figure}
\begin{center}
\includegraphics[width=\columnwidth]{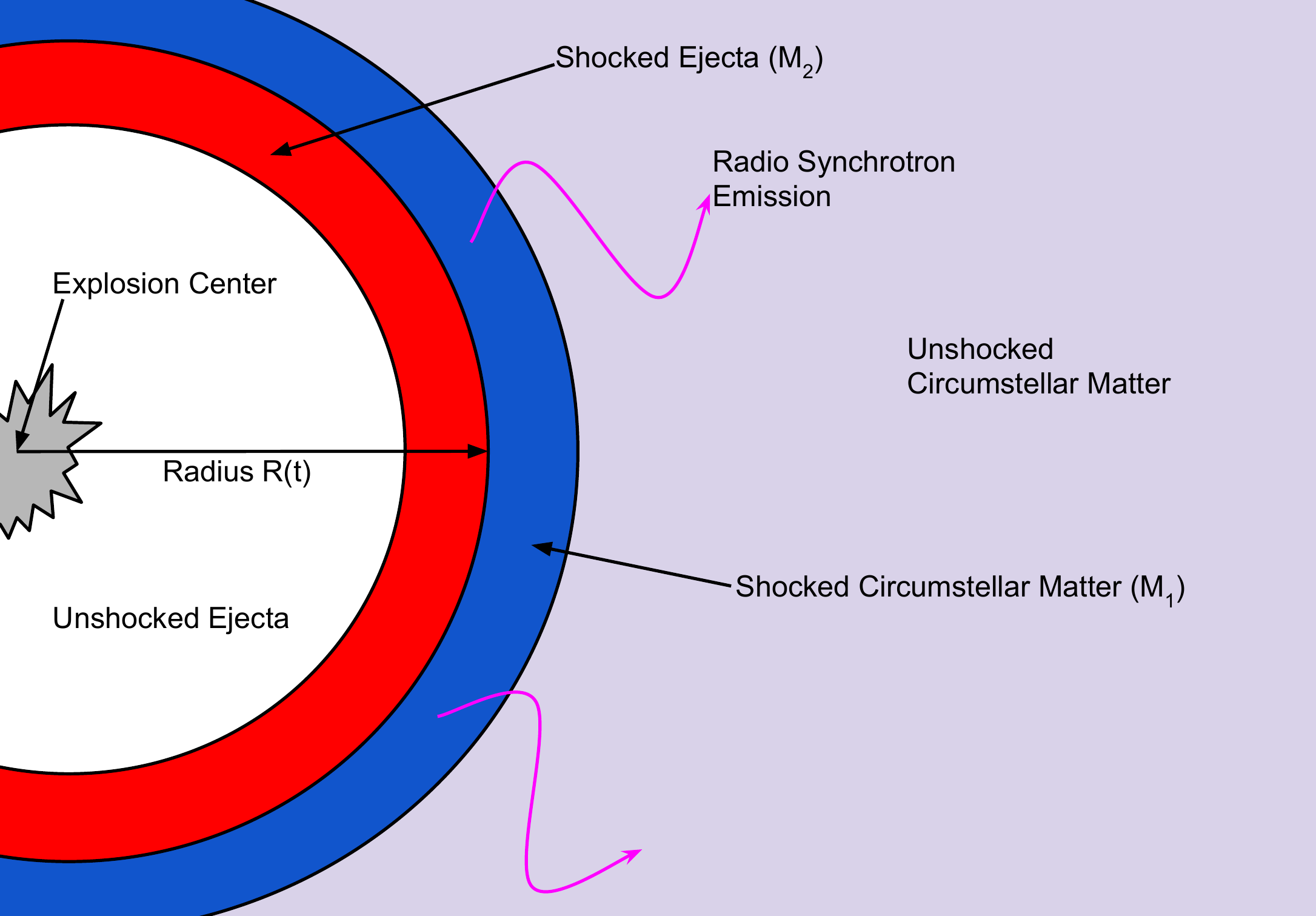}
\caption{Schematic representation of the model used to predict
the size and flux density evolution of young supernova remnants. The ejecta
and circumstellar matter are heated by shocks at the interaction. A fraction
of the thermal energy goes to magnetic fields and accelerated electrons.
The synchrotron emission from these electrons drive the flux density evolution.}
\label{fig:schematic}
\end{center}
\end{figure}

\subsection{Initial Conditions}
We consider a scenario (see Figure \ref{fig:schematic}) where the ejected
mass interacts with circumstellar
matter at a radius $R(t)$ \citep{2011ApJ...729...57C}. Following
\citet{1982ApJ...258..790C}, we label the mass of the shocked circumstellar
matter as $M_1$, and the shocked ejected mass as $M_2$. The density inside
the contact discontinuity is $\rho_{sn}$ and the density outside is
$\rho_{cs}$. The circumstellar density profile is different in the SD
and DD cases. In the SD case, the density is shaped by the mass loss
$(\dot{M})$ from the wind (with velocity $v_{\rm w}$). This can happen
in a various ways, such as loss from the outer Lagrange point or the
winds driven by accretion on to the degenerate companion. In the DD case,
where neither star has appreciable winds, the circumstellar environment
is essentially provided by the local density that remains mostly unaltered
by the binary and is assumed constant for this simple model.

We express the
pre-explosion circumstellar density ($\rho_{cs}$) at a distance $r$ as, 
\begin{equation}
\rho_{cs} \propto r^{-s}.
\end{equation}
In the above equation, the power law index $s$ is $2$ for the SD case and $0$
for the DD case. The presence of nova shells do not alter the situation.
The age of the remnant under consideration is approximately 150 years
\citep{2008MNRAS.387L..54G} and this is much larger than the time it would take
to sweep up the distance between individual shells, which is between 1 and
10 years. Thus, the granularity presented by the shells does not matter in the
long term evolution of the size and flux.

We assume that the fastest moving
ejecta has a power law density profile, 
\begin{equation}
\rho_{sn} \propto v^{-n} t^{-3} \propto r^{-n}t^{n-3}.
\end{equation}
This substitution is allowed for the ejecta in homologous expansion which has not yet
interacted with anything. This allows
us to use $v \equiv \frac{r}{t}$. Note that only a small fraction of the matter
ejected by the supernova is at the very high velocities. The steepness of
this profile is controlled by the power law index $n$, which must be greater
than 5 for the total energy in the ejecta to be finite. \citet{1969ApJ...157..623C}
suggested profiles with $n = 7$ for an explosion of a high mass white dwarf.
\citet{1984ApJ...286..644N} used their W7 model to explain the early spectral
evolution of of Type Ia supernovae. Models like these often had steep ejecta
profiles, prompting some authors to consider exponential profiles
\citep{1998ApJ...497..807D}.

\subsection{Blastwave Dynamics}
\label{blastwave_dynamics}
Here we consider the scaling relations of different parameters
depending on the circumstellar density profile and the explosion profile.
We begin by finding the mass of the shocked circumstellar matter, $M_1$ as
\begin{equation}
M_1 \propto \int_0^R{\rho_{cs} r^2}dr \propto R^{3-s}.
\end{equation}
This is simply the mass enclosed in the spherical region that has been hollowed
out by the explosion. By evaluating the integral we find $M_1$'s dependence
on radius. Next we find the amount of shocked ejected mass $M_2$ as
\begin{equation}
M_2 \propto \int_R^{\infty}{\rho_{sn} r^2}dr \propto t^{n-3}R^{3-n}.
\end{equation}
This is the mass that has already interacted with the circumstellar medium
and has been slowed down. In the same fashion as before we evaluate the integral to
find how $M_2$ depends on radius.

The next useful quantity to evaluate is the pressure. The pressure
that the shocked circumstellar mater exerts $P_1$, and the pressure
of the ejected mass $P_2$, compete to decelerate or accelerate the expansion
of the remnant. The pressure provided by the flux of momentum
brought in by the matter reaching the contact discontinuity is
proportional to the density times the square of the velocity. 
\begin{equation}
P_1 \propto \rho_{cs} (R')^2 \propto t^{-2}R^{2-s}
\end{equation}
\begin{equation}
P_2 \propto \rho_{sn} (R')^2 \propto t^{n-5}R^{2-n}
\end{equation}
These two equations are simplified  by substituting our expressions
for the different densities. We also know the shell is decelerating as
it interacts with the circumstellar material \citep{1982ApJ...259..302C}.
This deceleration is proportional to the difference in pressure times
the area of the shell, so
\begin{equation}
\label{eqn:mass_pressure}(M_1 + M_2)R''(t) \propto R^2(P_2-P_1).
\end{equation}
We use this to see how radius scales with time. 
\begin{equation}
(M_1 + M_2)  \frac{R}{t^2}  \propto R^2(P_2 - P_1)
\end{equation}

Therefore, after plugging in our previous scaling relationships
for $M_1, M_2, P_1, \rm{and} P_2$ we see that,
\begin{equation}
R \propto t^{\frac{n-3}{n-s}}.
\end{equation}
We call the power index of this equation $m$ from now on. So,
\begin{equation}
\label{eqn:radius-time} R \propto t^m,
\end{equation}
where
\begin{equation}
\label{eqn:m} m \equiv \frac{n-3}{n-s}.
\end{equation}

\subsection{Magnetic Fields and Particle Acceleration}
Now we use these scaling relations to figure out how the thermal energy,
magnetic field, number of accelerated electrons, and finally the flux scale with time.
We find the thermal energy by evaluating the kinetic energy lost during the
decelerated expansion.
\begin{equation}
E_{th} \propto M_1 (R')^2 \propto t^{\frac{(n-5)(3-s)}{n-s}}
\end{equation}
Note, that in both the SD and DD cases, as long as $t$ is less than the Sedov time, bulk
of the energy remains locked up in the kinetic energy of the ejecta. The
thermal energy $E$ is lesser than $E_0$ and steadily increasing during
this phase. As more and more gas is shock heated by the circumstellar
interaction, a fraction of this energy is made available for magnetic
field amplification and cosmic ray acceleration. This is what drives
the radio lightcurves of late time supernovae \citep{1982ApJ...258..790C}
and young remnants \citep{1984MNRAS.207..745C}.

Considering magnetic fields of average strength $B$, produced by turbulent amplification
at shocks, total magnetic energy scales as,
\begin{equation}
 E_B\propto B^2 R^3 .
\end{equation}
Following \citet{1982ApJ...258..790C} we consider, that a fraction
of the thermal energy goes into producing magnetic fields. Therefore, the
magnetic field scales as,
\begin{equation}
B \propto (E_{th}R^{-3})^{1/2} \propto t^{\frac{s(5-n) - 6}{2(n-s)}}.
\end{equation}

We consider a shock accelerated electron distribution where the number density
of energetic electrons is given by $N_0 E^{-p}dEdV$. Here $N_0$ is the
normalization of the spectrum of accelerated electron and $p$ is the power
law index of the same spectrum. We assume that this distribution extends
from $\gamma_m m_ec^2$ to infinity, filling a fraction of the spherical
volume of radius $R$. Therefore the total energy in accelerated electrons scales like,
\begin{equation}
E_e\propto N_0 R^3.
\end{equation}
Assuming that this represents a fraction of the total thermal
energy, we have,
\begin{equation}
N_0 \propto E_{th}R^{-3} \propto t^{\frac{s(5-n) - 6}{(n-s)}}.
\end{equation}

\subsection{Synchrotron Emission}
Early non-thermal emission from a supernova is often optically thick even
at radio radio frequencies, due to free-free or synchrotron self absorption.
As the optical thickness reduces with time, the flux density often rises.
The peaking of the radio light curve can take days to months depending upon
the circumstellar density and expansion velocity. However, late time radio
supernovae display optically thin spectra \citep{1982ApJ...258..790C}. Given
that the young remnant phase follows after the late supernova phase, we can
safely assume that the radio emission $(F_\nu)$ at a frequency $\nu \sim 1$ GHz is
reasonably approximated by an optically thin spectra.

Following \citet{1979rpa..book.....R} and \citep{1982ApJ...258..790C}
\begin{equation}
\label{eqn:flux}F_\nu=\frac{4 \pi f R^3}{3 D^2} c_5 N_0 B^{(p+1)/2} \left(\frac{\nu}{2c_1}\right)^{-(p-1)/2},
\end{equation}
where $c_1$ and $c_5$ are constants \citep{1970ranp.book.....P} and
$D$ is distance to source. Since we know $R$, $N_0$ and $B$ as functions
of time, we can calculate how radio flux density scales with time,
\begin{equation}
F_\nu \propto R^3 N_0 B^{7/4} \propto t^{\frac{3(-54+n(8-5s)+25s)}{8(n-s)}}.
\end{equation}
We will call the power index of time in the above equation $\beta$. So,
\begin{equation}
\label{eqn:flux-beta}F_\nu \propto t^{\beta}, 
\end{equation}
where, 
\begin{equation}
\label{eqn:beta}
\beta \equiv \frac{3(-54+n(8-5s)+25s)}{8(n-s)}.
\end{equation}

When we inspect the cases for SD or DD scenarios we see a striking
difference in how the flux scales with time.
\begin{equation}
F_{\nu} \propto 
\begin{cases}
t^{-\frac{3(2+n)}{4(n-3)}} & \text{for SD (s=2),}
\\
t^{3-\frac{45}{4(n-3)}} &\text{for DD (s=0).}
\end{cases}
\end{equation}
We note that in the SD case flux decreases with time , but in the DD case flux
can increase for explosions with steep ejecta profiles. This is a stark qualitative
distinction.

\begin{figure}
\begin{center}
\includegraphics[width=\columnwidth]{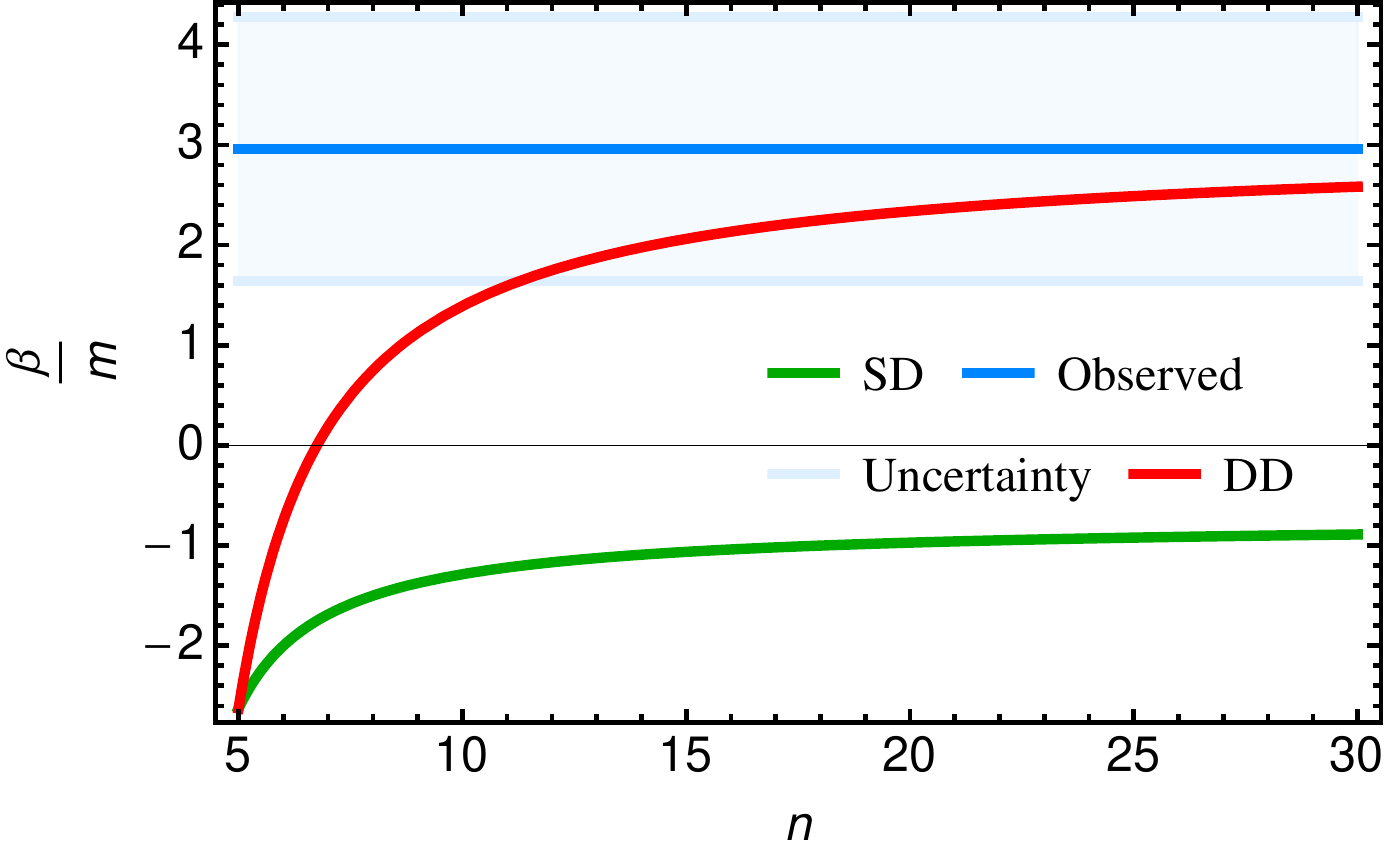}
\caption{Value of the Surface Brightness Index ($\frac{\beta}{m}$, which is equal to
$\frac{\dot{F}/F}{\dot{R}/R}$) observed from SNR G1.9 + 0.3 compared with the
modeled relationship between $\frac{\beta}{m}$ and $n$ for both the SD and DD
cases (Equation \ref{eqn:beta_m}). The plotted error margin is 2$\sigma$. Note
that the prediction from the DD case enters the allowed region, while that from
the SD case does not. We can use this to select the DD scenario and reject
the SD scenario.}
\label{fig:finding_n}
\end{center}
\end{figure}

\section{Surface Brightness Index from Chandra X-Ray Observations}
Recent X-ray observations from Chandra \citep{2008ApJ...680L..41R, 2014ApJ...790L..18B}
tell us that SNR G1.9+0.3 is expanding and has increasing flux.
\citet{2014ApJ...790L..18B} found that the flux increases at a rate of,
$\dot{F}/{F}=1.9 \pm 0.4{\rm \% \; yr^{-1}}$.
\citet{2008ApJ...680L..41R} measured that the supernova remnant is expanding
at a rate $\dot{R}/{R} = 0.642 \pm 0.049{\rm \% \; yr^{-1}}$.

We know from Equation \ref{eqn:radius-time} that
$\dot{R}/{R} = m/t$.
We can see from Equation \ref{eqn:flux-beta} that
$\dot{F}/{F} = \beta / t$. The right hand sides of
two relationships are derived from a theoretical standpoint. Their left hand sides
can be determined experimentally as listed above.
We can therefore eliminate the age of the remnant and solve for $\frac{\beta}{m}$.
This will help us decide the correct circumstellar density profile and allowed
values of the explosion index $n$, which can explain the evolution of the remnant.
Therefore, from observations,
\begin{equation}
\frac{\beta}{m} \equiv \frac{\dot{F}/F}{\dot{R}/R} = 2.96 \pm 0.66.
\end{equation}
We name this ratio, the Surface Brightness Index, since it relates flux
and size evolution. It is a dimensionless number
which measures how the brightness of the remnant evolves. Since it does not
explicitly depend on the age, it can be determined from observations even
when the date of explosion is unknown.

The Brightness Index can also be computed from theory.
Using the Equations \ref{eqn:m} and \ref{eqn:beta} for $m$ and $\beta$ we can
write from theory,
\begin{equation}
\frac{\beta}{m} = 
\begin{cases}
-\frac{3}{4}\times \frac{n^2-4}{(n-3)^2} & \text{for SD (s=2),}
\\
\frac{3}{4}\times \frac{n(4n-27)}{(n-3)^2} &\text{for DD (s=0).}
\end{cases} \label{eqn:beta_m}
\end{equation}
For an expanding remnant, a negative Brightness Index represents a declining
flux while a positive one denotes a rising flux. The two special cases
of $\frac{\beta}{m}=0$ and $2$ represent a constant flux and a constant
surface brightness respectively.

In Figure \ref{fig:finding_n} we graph the observed values of $\frac{\beta}{m}$ (and
its range of uncertainty) vs. $n$ (using the relationships found in Equation
\ref{eqn:beta_m}). This allows us to see which model is acceptable and what value of $n$
puts the preferred model in the observed band. Just from the equations we can see
that $\frac{\beta}{m}$ will always be negative in the SD case, so it will never
yield the observed result of simultaneously increasing flux and size. Thus a SD
solution only predicts decreasing flux, which we know to be untrue from the
observed data. After selecting the $s=0$ case based on observations, we can find our
allowed range of $n$. Using the chosen values of $s$ and $n$ we determine
the age of the remnant in the next section.

\section{Age Estimates}
Having selected the DD explanation $(s=0)$, based on observations, we now have
to pick a value of $n$.
Many authors, including \citet{1982ApJ...259L..85C}, have used a power law
profile with $n=7$ following \citet{1969ApJ...157..623C}. However, based on the
recent X-ray observations \citep{2014ApJ...790L..18B} and comparison with
our models in Figure \ref{fig:finding_n} we notice that we needed a steeper ejecta
profile governed by a larger value for $n$. From Figure \ref{fig:finding_n} we
notice that only models with values of $n \gtrsim 11.5$ would explain the data.
We chose 12 as our fiducial value of $n$ in the rest of this work. 

\subsection{DD Case}
Now we determine the age of the remnant using our value of $n=12$ and $s=0$,
for the DD case. To this end, we first find, 
\begin{equation}
m=\frac{n-3}{n} = \frac{3}{4}.
\end{equation}
We also know that,
\begin{equation}
\frac{\dot{R}}{R} = \frac{m}{t} = \frac{3}{4t} = 0.642 \pm 0.049 {\rm \% \; yr^{-1}}
\end{equation}
Solving for $t$ we have,
\begin{equation}
t=116.8 \pm 8.9 {\rm \; yr},
\end{equation}
and taking into account that this data was obtained in 2008, we
can determine when the supernova exploded. So from this estimate the
supernova occurred in $1892 \pm 9$ years.

We can also estimate the age
using our equation for $\beta$ and the change in flux over time.
We know,
\begin{equation}
\frac{\dot{F}}{F} = \frac{\beta}{t} = \frac{7}{4t} = 1.9 \pm 0.4 {\rm \% \; yr^{-1}}.
\end{equation}
Solving for $t$ in this case we have,
\begin{equation}
t=92.1 \pm 19.37 {\rm \; yr},
\end{equation}
So from this estimate the supernova occurred in $1916 \pm 19$ years.

The weighted mean of these two ages is  $109 \pm 9$ years, which gives an
explosion date of around $1899 \pm 9$. This is within the upper limit  of 150 years
proposed by \citet{2008MNRAS.387L..54G}.

\subsection{SD Case}
Even though the SD case was shown to be inapplicable to SNR G1.9.0.3, for a
consistency check we will now determine the age in the SD scenario using $s=2$.
So to start once again we find $m$ for this case.
\begin{equation}
m=\frac{n-3}{n-2} = \frac{9}{10}
\end{equation}
In the same fashion as above we will first solve for the age using the value of
\begin{equation}
\frac{\dot{R}}{R} = \frac{m}{t} = \frac{9}{10t} = 0.642 \pm 0.049 {\rm \% \; yr^{-1}}.
\end{equation}
So,
\begin{equation}
t=140.19 \pm 10.69 {\rm \; yr}.
\end{equation}

Now using the relationship for how flux changes over time, we will solve for $t$ again.
\begin{equation}
\frac{\dot{F}}{F} = \frac{\beta}{t} = \frac{-7}{6t} = 1.9 \pm 0.4 {\rm \% \; yr^{-1}},
\end{equation}
Where $\beta = -\frac{7}{6}$.
Thus,
\begin{equation}
t=-61.40 \pm 12.84 {\rm \; yr}.
\end{equation}
This points to an explosion date in the future,
which is absurd. This result is not physically plausible and merely shows
again that the SD case cannot incorporate a rising flux.

\section{Flux and Size Evolution}
Having picked the preferred scenario based on scaling relations, we can now
explicitly compute the flux and size evolution.
When $n=12$ the following equations describe the progression of the supernova
over time for the DD scenario. We start by looking again at what the circumstellar
density is,
\begin{equation}
\rho_{cs} = \rho_0.
\end{equation}

Next we enumerate the density of the supernova ejecta. We assume a broken power
law profile where the slow part has a constant density and
the fast part has a power law profile with $n=12$. The profile is determined by the
constant density and the velocity at the point of change.
These two values are completely determined by the initial energy and
the initial mass of the supernova ejecta. Here $E_0$ is the initial energy,
$v$ is the change over velocity, and $M_0$ is the initial mass.
\begin{equation}
\label{eqn:rho_sn}\rho_{sn} = \frac{13.0 E_0^{9/2}}{M_0^{7/2}t^3v^{12}}
\end{equation}

We can compare this with the initial setup and see that
the scaling is the same for $s=0,$ and $n=12$. Next we can find the
masses of the shocked circumstellar matter $(M_1)$, and the ejected
mass $(M_2)$. These were found by integrating the respective densities
over the relevant volumes.  
\begin{equation}
M_1 = \frac{4}{3}\pi \rho_0 R^3 
\end{equation}
\begin{equation}
M_2 = \frac{18.2E_0^{9/2}t^9}{M_0^{7/2}R^9}
\end{equation}
We again can see that these agree with the predicted scalings in
Section \ref{blastwave_dynamics}.

Next the two pressures are found by
multiplying the flux of momentum trying to cross the contact discontinuity.
$P_1$ is the pressure of the shocked cirucumstellar matter,
\begin{equation}
P_1 = \frac{0.640\rho_0}{ M_0^{2}} \left( \frac{\rho_0}{E_0^{9/2}M_0^{17/2}} \right)^{-1/6} t^{-1/2}.
\end{equation}
$P_2$ is the pressure of the ejected mass that collides with the shocked
circumstellar matter. Therefore, 
\begin{equation}
P_2 = \frac{0.427\rho_0}{M_0^2} \left( \frac{\rho_0}{E_0^{9/2}M_0^{17/2}} \right)^{-1/6} t^{-1/2}.
\end{equation}

We can then use Equation \ref{eqn:mass_pressure} to solve for the radius,
which comes out as ,
\begin{equation}
R(t)=\frac{1.07 }{M_0} \left( \frac{\rho_0}{E_0^{9/2}M_0^{17/2}}\right)^{-1/12}t^{3/4}.
\end{equation}

As stated above, thermal energy is less than the initial energy, but increasing
during this phase of the supernova. The kinetic energy lost due to the interaction
with the circumstellar medium is the thermal energy,
\begin{equation}
E_{th} = \frac{2.56 E_0^{15/8} \rho_0^{7/12}t^{7/4}}{M_0^{35/24}}.
\end{equation}

We consider a fraction, $f$, of the total volume, to be filled with amplified
magnetic fields.
So the energy in the magnetic field is,
\begin{equation}
E_B= \frac{0.202B^2 E_0^{9/8} f t^{9/4}}{M_0^{7/8}\rho_0^{1/4}}.
\end{equation}
Assuming that a fraction $(\epsilon_B)$ of the thermal energy goes into producing
this magnetic field, we get
\begin{equation}
B=\frac{3.56 E_0^{3/8}\left(\frac{\epsilon_B}{f}\right)^{1/2} \rho_0^{5/12}}{M_0^{7/24}t^{1/4}}.
\end{equation}

Similarly, to find the energy in the accelerated electrons, the
number density of electrons considered is $N_0 E^{-p}dEdV$
extending from $\gamma_m m_ec^2$ to infinity. These electrons are assumed to fill a fraction
$f$ of the spherical remnant with radius $R$. So the energy in the accelerated
electrons is,
\begin{equation}
E_e = \frac{5.08 E_0^{9/8}f (\gamma_m m_ec^2)^{2-p}N_0t^{9/4}}{M_0^{7/8}(p-2)\rho_0^{1/4}}.
\end{equation}
Assuming this is a fraction $(\epsilon_e)$ of the total thermal energy we have,
\begin{equation}
N_0 = \frac{0.503 E_0^{3/4}\epsilon_e(\gamma_m m_ec^2)^{p-2}(p-2)\rho_0^{5/6}}{fM_0^{7/12}t^{1/2}}.
\end{equation}

Using Equation \ref{eqn:flux} and assuming $\epsilon_e=\epsilon_B=0.01$ we have,
\begin{equation}
F_\nu = \frac{(8.260\times 10^{-6})c_5 E_0^{81/32}\left(\frac{c_1}{\nu}\right)^{3/4} (\rho_0t)^{21/16}}{D^2M_0^{63/32}}.
\end{equation}
The flux depends directly on the initial energy and inversely on the
distance to the supernova. It also increases with time and the initial
density of the explosion, in a limited fashion.

\section{Predictions for Observed Flux and Size} \label{predictions}
Here we recast the equations from the previous section in units which can be used
to conveniently predict fluxes and angular diameters, as determined by radio observations.

\subsection{DD Case}
First, the equation for the evolution of flux gives us,
\begin{align}
F_\nu = & 46.6 \left(\frac{t}{100 \rm yr}\right)^{1.31} \left(\frac{n_0}{\rm atom/cc}\right)^{1.31} \left(\frac{E_0}{10^{51} \rm ergs}\right)^{2.53} \nonumber \\ 
& \times \left(\frac{\nu}{\rm GHz}\right)^{-0.75} \left(\frac{D}{10\rm kpc}\right)^{-2} \left(\frac{M_0}{1.4M_{\odot}}\right)^{-1.97} \rm mJy.\label{flux_prediction}
\end{align}
This equation shows that the flux depends most strongly on the initial
energy of the explosion and rises with time.

Next the angular diameter, $\theta \equiv \frac{2R}{D}$, is given as
\begin{align}
\theta = & 42.7" \left(\frac{t}{100 \rm yr}\right)^{0.75} \left(\frac{E_0}{10^{51} \rm ergs}\right)^{0.38} \left(\frac{M_0}{1.4M_{\odot}}\right)^{-0.29} \nonumber \\
& \times \left(\frac{n_0}{\rm atom/cc}\right)^{-0.08} \left(\frac{D}{10\rm kpc}\right)^{-1}.\label{angular_prediction}
\end{align}
So the angular diameter is most strongly affected by the distance
from the observer to the SNR and the time since the explosion. 

\begin{figure}
\begin{center}
\includegraphics[width=\columnwidth]{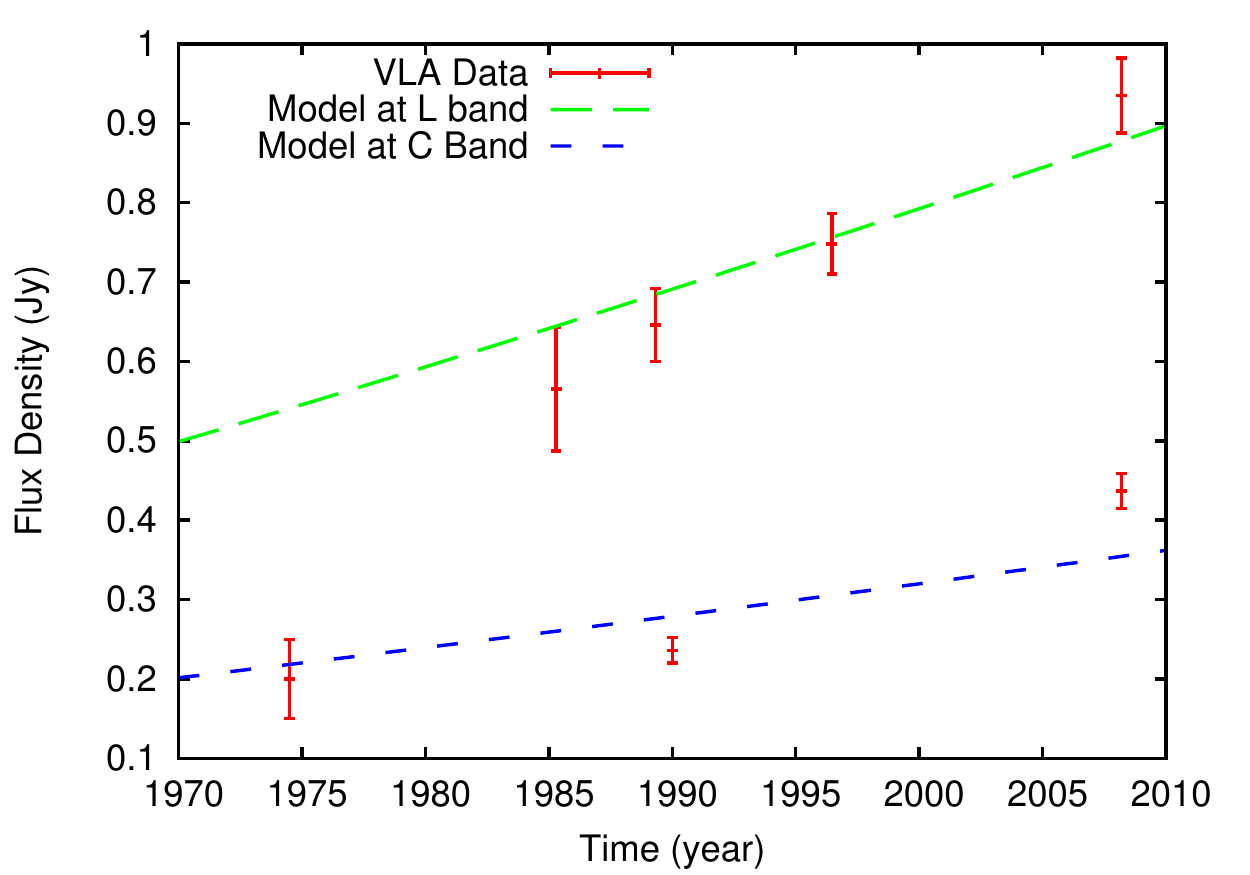}
\caption{Flux densities from VLA radio observations of SNR G1.9+0.3 at L and
C Bands compared with the best fit from the DD scenario (Equation \ref{flux_prediction}).
The data and the favored DD model, both show a rise. In contrast
the disfavored SD model (Equation \ref{flux_prediction_SD}) would have predicted a decaying lightcurve.}
\label{fig:radio_observations}
\end{center}
\end{figure}

\subsection{SD Case}
Here we provide the corresponding versions of these equations, in
convenient units, for the SD case. These are not used for SNR G1.9+0.3 but
are provided for reference. First, the flux can be expressed as
\begin{align}
F_\nu = & 33.1 \left(\frac{t}{100 \rm yr}\right)^{-1.05} \left(\frac{\dot{M}}{10^{-7}M_{\odot}/\rm yr}\right)^{1.58} \nonumber \\ 
& \times \left(\frac{E_0}{10^{51} \rm ergs}\right)^{1.35} \left(\frac{\nu}{\rm GHz}\right)^{-0.75} \left(\frac{D}{10\rm kpc}\right)^{-2} \nonumber \\
& \times \left(\frac{M_0}{1.4M_{\odot}}\right)^{-1.05} \left(\frac{v_w}{100\rm km/s}\right)^{-1.58} \rm mJy.\label{flux_prediction_SD}
\end{align}
We can see from this equation that flux in this case strongly depends on the
distance to the object, $D$. The rate of mass loss, $\dot{M}$ and the velocity
of the wind, $v_w$ also affect the flux.

Next the angular diameter can be written as
\begin{align}
\theta = & 63.9" \left(\frac{t}{100 \rm yr}\right)^{0.9} \left(\frac{E_0}{10^{51} \rm ergs}\right)^{0.45} \left(\frac{M_0}{1.4M_{\odot}}\right)^{-0.35} \nonumber \\
& \times \left(\frac{\dot{M}}{10^{-7}M_{\odot}/\rm yr}\right)^{-0.1} \left(\frac{v_w}{100\rm km/s}\right)^{0.1} \left(\frac{D}{10\rm kpc}\right)^{-1}.
\end{align}
The angular diameter depends on the distance from the explosion, and the
time since the explosion. There is also a large direct dependence on the
initial energy.

We provide results for both the DD and SD scenarios so that they can be 
compared with future observations of other remnants to see which one more
accurately models the observed evolution.

\section{Comparison with VLA Radio Observations}
We can use the recent radio observations to estimate the size and flux of the
supernova remnant. Equations \ref{flux_prediction} and \ref{angular_prediction} let us
find how they depend on external density and initial energy. From Figure 2 in
\citet{2008MNRAS.387L..54G} that depicts the azimuthally averaged radial profile
of the radio emission in 2008, we inferred a mean emission weighted radius of
$\frac{\theta}{2}=34.5''$ at an age of $t=109$ yr.
This observed size can be substituted into
Equation \ref{angular_prediction} recast as,
\begin{equation}
\theta= 53.6'' E_{51}^{3/8} n_0^{-1/12},
\end{equation}
for the distance to the particular remnant and its age.
In the above equation the only two unknowns are
$n_0$ (in units of atoms$/$cc) and $E_{51}$ (in units of $10^{51}$ ergs)
as $\theta$ is known from observations.

From the Section \ref{predictions} we can use Equation \ref{flux_prediction},
with a value of flux ($F_0$) for observations at $\sim 1$ GHz. This can then
be compared to the observed value from \citet{2008MNRAS.387L..54G}.
We also use a distance $D = 8.5 \rm kpc$. We chose this
as the distance to the supernova because we assume it is near the Galactic center.
Time of the observations is measured from our fiducial explosion year of 1899.
Mass of the explosion is assumed to be close to a Chandrasekhar mass. First we fit
(See Figure \ref{fig:radio_observations}) the radio observation with,
\begin{equation}
F_{\nu}(t,\nu) = F_0 \left(\frac{(t-1899)}{100 \rm yr}\right)^{21/16}\left(\frac{\nu}{1 \rm GHz}\right)^{-3/4} \rm Jy,
\end{equation}
to find $F_0$, the flux density at age 100 years observed at 1 GHz. We did find
that $F_0 = 1.03 \pm 0.05 \rm Jy$. Next we can plug this information into,
\begin{equation}
F_{0} = 64.5 E_{51}^{81/32} n_0^{21/16} \rm mJy.
\end{equation}

Substituting the observed size and flux into the Equations \ref{flux_prediction}
and \ref{angular_prediction}, we can therefore 
solve for the density, $(n_0)$, and the initial energy, $(E_{51})$.
We found that the values, $n_0 = 1.8 {\rm\; atom/cc}$ and $E_{0}=2.2\times 10^{51} \rm ergs$,
when used in Equations \ref{flux_prediction} and \ref{angular_prediction}, reproduce the
observed flux and size evolution in the radio. We also note that these are reasonable values
to expect for the density and initial energy. Note that these values should be seen as
consistency checks rather than determinations of the density and energy, because of
systematic uncertainties introduced by unknowns like $\epsilon_e$ and $\epsilon_B$.

\section{Effects of Electron Cooling}
The effects of electron cooling on broadband spectrum may become apparent when
observing the spectrum in both the radio and the x-ray. Magnetic fields that 
permeate the SNR cause the electrons to lose energy, and cool down. This produces
emission that can be described by a power law, 
\begin{equation}
F_{\nu} \propto \nu^{\alpha}.
\end{equation}
However, the radio and x-ray emission may not be explained by the same power law.

\subsection{Spectral model}
Above some critical Lorentz factor $(\gamma_c)$ the electrons have lost enough
energy that slope of the power law changes. According to \citet{1999PhR...314..575P}
the slope before the critical Lorentz factor, possibly at the radio frequencies, is 
\begin{equation}
{\alpha}=\frac{1-p}{2},
\end{equation}
where we chose $p$ to be $2.5$. Once the electrons' Lorentz factor is above
$\gamma_c$ the slope changes to \citep{1999PhR...314..575P}, 
\begin{equation}
{\alpha}=\frac{-p}{2}.
\end{equation}

The Lorentz factor, of radiating electrons, is related to a corresponding
frequency $(\nu_c)$, of emitted photons. Following
\citet{1979rpa..book.....R}, these can be related as,
\begin{equation}
\nu(\gamma)=\gamma^2\frac{q_eB}{2\pi m_e c},
\end{equation}
where $c$ is the speed of light, $B$ is the magnetic field, and $m_e$ mass of an
electron. The critical Lorentz factor, above which synchrotron losses dominate,
is given by \citep{1998ApJ...497L..17S},
\begin{equation}
\gamma_c = \frac{6\pi m_e c}{\sigma_T B^2t},
\end{equation}
where $\sigma_t$ is the Thomson cross-section, and $t$ is the age of the remnant.
This critical Lorentz factor comes out to be, $\gamma_c \sim 2 \times 10^5$
after evaluating
the expression for SNR G1.9+0.3. \citet{2011ApJ...729...57C} expressed the
relationship, between the critical frequency and Lorentz factor, as
\begin{equation}
\nu_c = \frac{18\pi m_e c q_e}{\sigma_T^2B^3t^2},
\end{equation}
where $q_e$ is the charge of an electron. Following this equation we found
that $\nu_c=1.3\times10^{14} \rm Hz$, so we expect that the value of
$\alpha$ should change at infrared frequencies. We checked our
estimates by increasing and decreasing the assumed density by a factor
of 10 to see if it would affect the break where the spectrum changes slope.
However, even after varying the density the break was still in the infrared.
Therefore we expect to see change in slope somewhere between the radio and x-ray
bands, probably in the infrared.

\begin{figure}
\begin{center}
\includegraphics[width=\columnwidth]{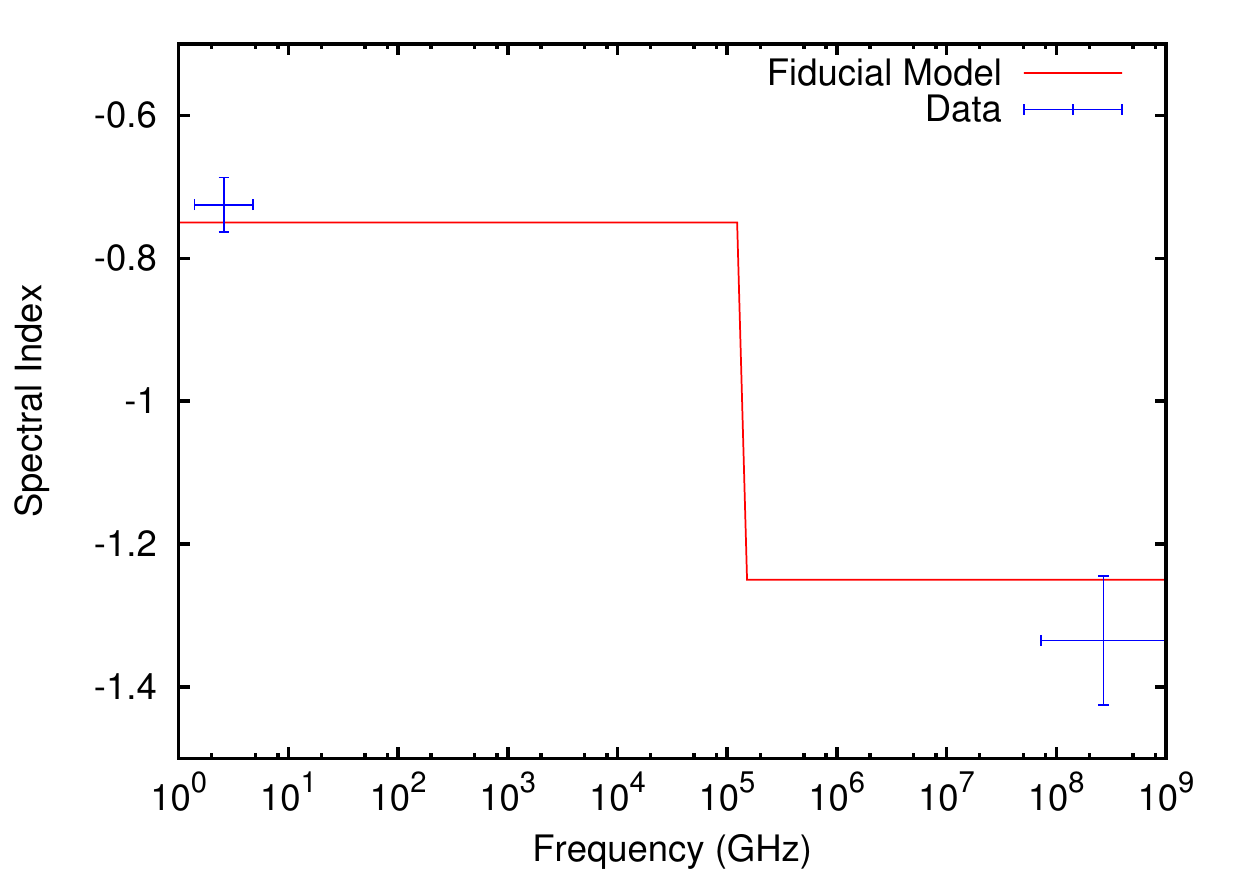}
\caption{This graph, of the spectral index predicted by
the fiducial model and that observed in the data, shows that the model can explain the
two different spectral indices at radio and X-ray frequencies. Note that this is not
a fit, but merely a graphical comparison of the predicted spectral indices with
observed ones.}
\label{fig:fiducial_vs_data}
\end{center}
\end{figure}

\subsection{Observed Spectral Indices}
The L and C band data from Figure \ref{fig:radio_observations} was used to
determine the spectral index at radio frequencies. L band corresponds to
frequencies near 4.8 GHz, while C band corresponds to 1.4 GHz in frequency.
We found that the spectral index was $\alpha_{\rm radio} = -0.725 \pm 0.091$

SNR G1.9+0.3 was observed with the Chandra X-Ray Observatory
by a team (PI: Kazimierz Borkowski) in Observation 12691 on May 9, 2011. The ACIS-S
chip was used for 184.0 ks. To use the data from the x-ray observations we first had
to extract the spectra out of the image. Then we imported this spectra into XSPEC
to further analyze it.  We used the tbabs absorption model and a simple power law emission
model to fit the data.
We argue against using the srcut model extending from radio to X-rays, because we
expect a synchrotron cooling break below the x-ray band. We therefore fit the radio
and X-ray data separately. The X-ray model was decided to be a good fit, by simulating
10,000 spectra where only $57\%$ of realizations were found better than the observed spectra.
We convert the photon index, as mostly used in X-ray analysis software, into the spectral
index to compare with radio observations. We found that $\alpha_{\rm Xray} = -1.335 \pm 0.045$.

Finally, in Figure \ref{fig:fiducial_vs_data} we plot the observed values of the spectral
index and compare it with the predictions from our fiducial model. Note, that this is not
a fit, but merely a comparison to show that our model naturally predicts the observed steepening
in the spectral index from the radio to the x-rays.

\section{Discussions}
We have shown that circumstellar interaction in young supernova remnants is an
useful tool in trying to discern progenitors of thermonuclear supernovae.
We have developed a new diagnostic, the Surface Brightness Index, relating
flux and size evolution of remnants.
In particular, we favor a DD scenario for SNR G1.9+0.3.
Based on application of our models to SNR G1.9+0.3 we present here a prescription
to evaluate other young supernova remnants that are observed in x-ray and/or radio.
These recommendations would be helpful in deducing the nature and surroundings of
a Type Ia supernova.
\begin{itemize}
\item Observe the flux density and size of the remnant to determine how they change in time.
\item Re-construct Figure \ref{fig:finding_n} and plot the observed ratio of the fractional changes in flux and radius.
\item Compare this observed ratio with theoretical predictions for the Surface Brightness Index. 
\item Rule out either SD or DD scenarios. 
\item Use appropriate equations from Section \ref{predictions}, to determine explosion energy and circumstellar density.
\end{itemize}

Type Ia supernovae are responsible for much of the heavy elements in the
universe and of fundamental importance as distance indicators in cosmology.
In such a situation, identification of their progenitors is a matter
of utmost concern. Opinion is divided mostly between SD and DD scenarios.
In this work we have shown that circumstellar interaction in young supernova
remnants can be a useful discriminator between these possibilities. We have
used this technique to demonstrate that the youngest Galactic supernova remnant
SNR G1.9+0.3 is likely the product of a DD progenitor system. Our model shows
that an SD scenario cannot produce a rising flux, whereas the DD case does. 
Our result shows that Type Ia supernovae can all have DD progenitors or a
combination of SD and DD populations. The scenario in which all progenitors
are SD is ruled out.

We suggest that further progress can be made by deep radio detections or tight
upper limits, thanks to the increased sensitivity of the upgraded VLA, of
historical Type Ia supernova in the local group.
Radio observations of nearby Type Ia supernovae within a year of explosion
\citep{2015arXiv151007662C} put tight constraints on SD progenitor scenarios.
Late observations will be particularly useful for constraining DD progenitor
scenarios since they predict rising flux densities.
Observations of SN 1885A
in M31, SN 1895B in NGC5253 and SN 1937C in IC4182 with $\sim \mu$Jy level
sensitivity will be especially useful. Future observations of SN 2011fe
are also important, even with the current upper limits (which puts
pressure on SD scenarios) as the DD scenario predicts a rising lightcurve
which may be detectable in the future.
In absence of pre-explosion progenitor detections, for Type Ia supernovae,
circumstellar interactions may provide the most important insights into
their hitherto elusive progenitors.

\section*{Acknowledgements}
We thank Alak Ray, Naveen Yadav, Roger Chevalier and Laura Chomiuk for discussions.
This work made use of radio observations from the NRAO VLA.
The National Radio Astronomy Observatory is a facility of the National
Science Foundation operated under cooperative agreement by Associated
Universities, Inc.
The scientific results reported in this article are based in part on
data obtained from the Chandra Data Archive.

\bibliography{biblio}

\end{document}